\documentclass[twocolumn,nofootinbib]{revtex4-1}
\usepackage{graphicx}
\usepackage{dcolumn}
\usepackage{bm}
\usepackage{amsmath}
\usepackage{amsfonts} 
\usepackage{latexsym}
\usepackage{bbm}
\usepackage{color}
\begin{document}


\title{Sphaleron And Critical bubble in a scale invariant model : ReAnalysis}

\author{Kaori Fuyuto$^{1}$}%
\email{fuyuto@th.phys.nagoya-u.ac.jp}
\author{Eibun Senaha$^{1,2}$}%
\email{senaha@ncu.edu.tw}
\affiliation{$^1$Department of Physics, Nagoya University, Nagoya 464-8602, Japan}
\affiliation{$^2$Department of Physics and Center for Mathematics and Theoretical Physics, National Central University, Taoyuan, 32001, Taiwan}
\bigskip

\date{\today}

\begin{abstract}
We revisit the electroweak phase transition and the critical bubble 
in the scale invariant two Higgs doublet model in the light of recent LHC data. 
Moreover, the sphaleron decoupling condition is newly evaluated in this model. 
The analysis is done by using the resumed finite-temperature one-loop effective potential. 
It is found that the 125 GeV Higgs boson inevitably 
leads to the strong first-order electroweak phase transition,
and the strength of which is always large enough to satisfy the sphaleron decoupling condition,
$v_N/T_N > 1.2$, where $T_N$ denotes a nucleation temperature and $v_N$ is the Higgs
vacuum expectation value at $T_N$.
In this model, even if the Higgs boson couplings to gauge bosons and fermions 
are similar to the standard model values, 
the signal strength of the Higgs decay to two photons is reduced by 10\%
and the triple Higgs boson coupling is enhanced by 82\% compared to the standard model prediction.  
\end{abstract}

\pacs{Valid PACS appear here}

\maketitle

\section{Introduction}
One of the observational facts that needs new physics beyond the standard model (SM) is the baryon asymmetry of the Universe (BAU) \cite{Ade:2013zuv},
\begin{align}
\frac{n_B}{s}=(8.59\pm0.11)\times 10^{-11}\hspace{0.5cm}({\rm Planck})
\end{align}
where $n_B~(s)$ denotes the baryon number (entropy) density. 
Although many mechanisms that can explain the observed value exist in the literature, 
electroweak baryogenesis (EWBG) \cite{ewbg} is the only scenario that is ripe for verification
by collider experiments, such as the Large Hadron Collider (LHC), 
and by low energy experiments, such as the electric dipole moments of the neutron, 
atoms and molecules. 
Since EWBG is intimately connected to Higgs physics,
the establishment of the Higgs sector plays an essential role in testing it, and
the discovery of the Higgs boson at the LHC in 2012 \cite{Aad:2012tfa,Chatrchyan:2012ufa}
is the first step toward the collider probe of EWBG.
Indeed, since the Higgs boson mass that is one of the relevant parameters 
for the electroweak phase transition (EWPT) has been measured with 0.2\% accuracy, 
$m_H=125.09\pm 0.21~({\rm stat.})\pm 0.11~({\rm syst.})$ GeV~\cite{Aad:2015zhl}, 
the feasible regions of EWBG have been narrowed down in various models~\cite{MSSM-EWBG_LHCtension}  . 
In upcoming experiments, such as the LHC Run-II and the High-Luminosity LHC~\cite{HL-LHC}, 
the Higgs boson couplings to the SM particles would be measured with better precision, and the international linear collider (ILC) \cite{Baer:2013cma} has the great capability of measuring the triple 
Higgs boson coupling, which may yield a decisive clue to the EWBG hypothesis.

In order for EWBG to be successful, the EWPT has to be strongly first order.
The properties of the EWPT are related not only to the Higgs boson mass and model parameters
but also to electroweak symmetry breaking mechanisms.
An interesting possibility is the so-called Coleman-Weinberg (CW) 
mechanism~\cite{Coleman:1973jx,Gildener:1976ih} 
in which quantum effects induce the symmetry breaking.
The scale invariant two Higgs doublet model (SI-2HDM)~\cite{Inoue:1979nn,Funakubo:1993jg,Takenaga:1993ux,Takenaga:1995ms, Lee:2012jn, Hill:2014mqa} 
is one of such examples, 
\footnote{The CW mechanism does not work in the SM 
since the top quark is too massive to give the stable vacuum.}
and the previous work \cite{Funakubo:1993jg} shows that the SI-2HDM can have 
the strong first-order EWPT. 
At the time of their analysis, however, 
the masses of the Higgs boson and top quark were not known.
Moreover, on the theoretical front,
neither a thermal resummation for the effective potential nor  
the evaluation of a baryon number preservation condition 
(also called a sphaleron decoupling condition) were conducted
in Ref.~\cite{Funakubo:1993jg}.

In this Letter, we update the analysis of the EWPT including the evaluation of bubble wall profiles,
and obtain the sphaleron decoupling condition by taking the recent LHC data into account.
In our study, we use the finite-temperature one-loop effective potential with daisy resummation. 
The phenomenological consequences of the sphaleron decoupling condition 
is also briefly discussed.
As studied in the previous works~\cite{Kanemura:2004ch, Fuyuto:2014yia},
we evaluate the deviations of the Higgs boson couplings from their SM values 
in the region where the strong first-order EWPT is achieved.

The paper is organized as follows. We give a quick review of the SI-2HDM in section~\ref{sec:Model},
and the Higgs boson couplings are presented in section \ref{sec:higgs}.
The sphaleron decoupling condition and the critical bubbles are discussed 
in section \ref{sec:sph_cb}. 
We show our results in section \ref{sec:Results}, 
and conclusions and discussions are given in section \ref{sec:conclusion}. 
%
%
%
%
%
%
%
%
%
%
%
%
%
%
%
%
%
%

\section{The model}\label{sec:Model}
The SI-2HDM is a minimal scale invariant extension of the SM 
by adding another Higgs doublet field.
The most general Higgs potential at the renormalizable level is given by  
\begin{align}
V_0=&\frac{\lambda_1}{2}(\Phi^{\dagger}_1\Phi_1)^2+\frac{\lambda_2}{2}(\Phi^{\dagger}_2\Phi_2)^2+\lambda_3(\Phi_1^{\dagger}\Phi_1)(\Phi_2^{\dagger}\Phi_2) \nonumber \\
&+\lambda_4(\Phi^{\dagger}_1\Phi_2)(\Phi^{\dagger}_2\Phi_1)
	+\bigg\{\frac{\lambda_5}{2}(\Phi^{\dagger}_1\Phi_2)^2 \nonumber \\
&+\lambda_6(\Phi^{\dagger}_1\Phi_1)(\Phi^{\dagger}_1\Phi_2)
	+\lambda_7(\Phi^{\dagger}_2\Phi_2)(\Phi^{\dagger}_1\Phi_2) +{\rm h.c.} \bigg\}.
\end{align}
After two Higgs doublets get vacuum expectation values (VEVs), they are cast into the form
\begin{align}
\Phi_i(x)=
\begin{pmatrix}
\phi^+_i(x)\\
\frac{1}{\sqrt{2}}\left(v_i+h_i(x)+ia_i(x)\right)
\end{pmatrix},\quad i=1,2,
\end{align}
where $v_1=v\cos\beta$ and $v_2=v\sin\beta$ with $0\le \beta \le \pi/2$, and $v\simeq 246$ GeV.
In order to avoid Higgs-mediated flavor changing neutral current (FCNC) processes at the tree level,
we impose a $Z_2$ symmetry ($\Phi_1\to -\Phi_1$, $\Phi_2\to \Phi_2$), 
which leads to $\lambda_6=\lambda_7=0$~\cite{Glashow:1976nt}.
The phase of $\lambda_5$ is removed by an appropriate field redefinition of the Higgs doublets,
so that CP is conserved. 

Following a method by Gildener and Weinberg~\cite{Gildener:1976ih}, 
we consider the EW symmetry breaking in a flat direction.
The tree-level effective potential takes the form
\begin{align}
V_0(\varphi_1,\varphi_2)=\frac{\lambda_1}{8}\varphi_1^4+\frac{\lambda_2}{8}\varphi^4_2+\frac{\lambda_{345}}{4}\varphi^2_1\varphi^2_2,
\end{align}
where $\varphi_1$ and $\varphi_2$ are the constant background fields of the two Higgs doublets. 

The tadpole conditions that are defined as the first derivatives of $V_0$ with respect to 
$\varphi_{1,2}$ give the following conditions:
\begin{align}
\lambda_{345}+\sqrt{\lambda_1\lambda_2}=0,\hspace{0.5cm}
\lambda_1v^4_1=\lambda_2v^4_2, \label{tadpole}
\end{align}
where $\lambda_{345}=\lambda_3+\lambda_4+\lambda_5$. 
With these conditions, it follows that $V_0(v_1,v_2)=0$. 
Moreover, since the mass matrix of $h_1$ and $h_2$ is written as
\begin{align}
{\cal M}^2_{\rm tree} = 
\begin{pmatrix}
\lambda_1v^2_1 & \lambda_{345} v_1 v_2\\
\lambda_{345} v_1 v_2 & \lambda_2 v_2^2
\end{pmatrix},
\end{align}
one finds ${\rm det}({\cal M}^2_{\rm tree}) =0$ using Eq.~(\ref{tadpole}). 
The appearance of the massless particle is the consequence of the classical scale invariance.  
We define $h$ and $H$ as the mass eigenstates of the CP-even Higgs bosons, 
which are obtained by 
\begin{align}
\begin{pmatrix}
h_1\\
h_2
\end{pmatrix}=
\begin{pmatrix}
\cos\alpha & -\sin\alpha\\
\sin\alpha & \cos\alpha
\end{pmatrix}
\begin{pmatrix}
H\\
h
\end{pmatrix}, 
%
%
%
%
\end{align}
where $-\pi/2 \leq \alpha \leq 0$. In the following discussion, $h$ is the SM-like Higgs boson
whose mass is zero at the tree level and is generated by the quantum corrections.
It can be proved that $\alpha=\beta-\pi/2$ at the tree level, and
consequently, the Higgs boson couplings to the gauge bosons and fermions 
are the same as those in the SM.

As mentioned above, $h$ becomes massive as the result of the radiative EW symmetry breaking.
The one-loop effective potential is~\cite{Coleman:1973jx,Jackiw:1974cv}
\begin{align}
V_1(\varphi)=\sum_in_i\frac{\bar{m}^4_i(\varphi)}{64\pi^2}\left(\log\frac{\bar{m}^2_i(\varphi)}{\bar{\mu}^2}-c_i \right),
\end{align}
where $\varphi=\sqrt{\varphi^2_1+\varphi^2_2}$ and $i=H,~A,~H^{\pm},~W^{\pm},~Z,~t,~b$, and 
$c_i=3/2~(5/6)$ for scalars and fermions (gauge bosons) and $\bar{\mu}$ denotes 
a renormalization scale. 
$A$ and $H^{\pm}$ are the physical CP-odd and charged Higgs bosons, respectively. 
$n_i$ are the degrees of freedom and the statistics of the particle $i$:
\begin{align}
n_H&=n_{A}=1,\quad n_{H^\pm}=2,\hspace{0.5cm}n_{W^{\pm}}=3\cdot2,\nonumber \\
n_{Z}&=3,\hspace{0.5cm} n_t=n_b=-12.
\end{align}
The field-dependent masses can be written as $\bar{m}^2_i=m^2_i\varphi^2/v^2$,
where $m_i$ are the corresponding masses in the vacuum, so that
$V_1(\varphi)$ is reduced to
\begin{align}
V_1(\varphi)=A\varphi^4+B\varphi^4\log\frac{\varphi^2}{\bar{\mu}^2},\label{V1}
\end{align}
with
\begin{align}
A=\sum_in_i\frac{m^4_i}{64\pi^2v^4}\left(\log\frac{m^2_i}{v^2}-c_i\right),\quad
B=\sum_in_i\frac{m^4_i}{64\pi^2v^4}.\label{A_B}
\end{align}
As can be seen from the tadpole condition of $V_1(\varphi)$,
we have a relationship between the scale of $v$ and the renormalization scale $\bar{\mu}$, 
i.e., $v^2=\bar{\mu}^2e^{-1/2-A/B}$, as the consequence of dimensional transmutation.
From Eqs.~(\ref{V1}) and (\ref{A_B}), 
it is easily checked that the vacuum energy becomes $V_1(v)=-Bv^4/2$, 
which implies that the electroweak symmetry is broken unless $B$ is negative. 
It should be noted that since $A$ and $B$ are the same order in the coupling, 
i.e., $\mathcal{O}(g^4)$, 
where $g$ collectively denotes the coupling constants
in this model, $A/B$ should be $\mathcal{O}(1)$, so the $\log (v^2/\bar{\mu}^2)\sim \mathcal{O}(1)$.
In other directions, however, $A$ may be $\mathcal{O}(g^2)$ and thus
$\log (v^2/\bar{\mu}^2)\sim1/g^2$, which may invalidate the perturbative calculation,
as advocated in Ref.~\cite{Gildener:1976ih}.

The mass of $h$ is obtained by taking the second derivative of $V_1(\varphi)$
and evaluating it at $\varphi=v$,
\begin{align}
m^2_h=\frac{\partial^2V_1(\varphi)}{\partial\varphi^2}\bigg|_{\varphi=v}=8Bv^2. \label{Higgs_mass}
\end{align}
We remark that thanks to the loop contributions from $H,~A$ and $H^{\pm}$, 
$B$ can be positive in contrast to the SM case, rendering $m^2_h$ positive. 
Interestingly, once $m_h=125~{\rm GeV}$ is fixed, 
the possible ranges of $m_H,~m_A$ and $m_{H^{\pm}}$ are restricted. 
In this Letter, we consider a case that $m_A=m_{H^{\pm}}$
in order to satisfy the constraint coming from the $\rho$ parameter~\cite{rho_para}.
Therefore, the heavy Higgs mass scales are specified by only two parameters.  
In what follows, $m_H$ and $m_A$ are chosen.

\section{Higgs boson couplings}\label{sec:higgs}

The Higgs boson couplings to gauge bosons and fermions normalized to the SM values are,
respectively, given by
\begin{align}
\kappa_V=\frac{g^{\rm SI\mathchar`-2HDM}_{hVV}}{g^{\rm SM}_{hVV}}, \hspace{1cm}
\kappa_f=\frac{g^{\rm SI\mathchar`-2HDM}_{hff}}{g^{\rm SM}_{hff}}, 
\end{align}
where $f = u, d, l.$ 
As discussed in section II, $\kappa_V=\kappa_f=1$ due to $\alpha=\beta-\pi/2$ at the tree level. 
Even in such a situation, the so-called nondecoupling effects may appear in the loop processes.
For instance, as pointed out in Ref.~\cite{Ginzburg:2002wt}, 
the $h\to \gamma\gamma$ mode may be significantly modified by the charged Higgs boson loop.
The Higgs signal strength of $h\to \gamma\gamma$ is defined as
\begin{align}
\mu_{\gamma\gamma}
&=\frac{\sigma(pp\to h)_{\rm SI\mathchar`-2HDM}{\rm Br}(h\to\gamma\gamma)_{\rm SI\mathchar`-2HDM}}{\sigma(pp\to h)_{\rm SM}{\rm Br}(h\to\gamma\gamma)_{\rm SM}}\nonumber\\
&\simeq \left|1+\frac{\mathcal{A}_{H^\pm}}{\mathcal{A}_{\rm SM}}\right|^2,
\end{align}
where $\mathcal{A}_{\rm SM}=-6.49$~\cite{McKeen:2012av} 
and $\mathcal{A}_{H^\pm}=-\tau_{H^\pm}\big(1-\tau_{H^\pm} f(\tau_{H^\pm})\big)$ 
with $\tau_{H^\pm}=4m_{H^\pm}^2/m_h^2$, and $f$ is a loop function defined
in Ref.~\cite{Gunion:1989we}.

The another nondecoupling effect may appear in the triple Higgs boson coupling.
The deviation of the triple Higgs boson coupling from its SM value is defined as
\begin{align}
\Delta \lambda_{hhh}=\frac{\lambda^{\rm SI\mathchar`-2HDM}_{hhh}-\lambda^{\rm SM}_{hhh}}{\lambda^{\rm SM}_{hhh}}.
\end{align}
In this analysis, we use the following expression as the SM prediction~\cite{hhh}
\begin{align}
\lambda^{\rm SM}_{hhh}=\frac{3m^2_{h}}{v}\left[1+\frac{9m^2_h}{32\pi^2v^2}+\sum_{i=W,Z,t,b}n_i\frac{m^4_i}{12\pi^2m^2_hv^2}\right].
\end{align}
Note that the dominant one-loop contribution comes from the top quark loop, 
which renders $\lambda_{hhh}$ smaller compared to the leading result.
In the SI-2HDM, the triple Higgs boson coupling to leading order
is simply expressed in terms of $m_h$ and $v$~\cite{Dermisek:2013pta}
\begin{align}
\hspace{1cm}\lambda^{\rm SI\mathchar`-2HDM}_{hhh}=
\frac{\partial^3V_1(\varphi)}{\partial\varphi^3}\bigg|_{\varphi=v}=40Bv = \frac{5m^2_h}{v}.
\end{align}
Unlike the ordinary 2HDM, the leading result in the SI-2HDM 
does not same as the leading one in the SM even in the case that $\beta-\alpha=\pi/2$,
which reflects the different origins of the electroweak symmetry breaking.

%
%
%
%
%
%
%
%
%
%
%
%
\section{Sphaleron decoupling condition and critical bubbles}\label{sec:sph_cb}
In EWBG, in order to preserve the generated BAU until today, the sphaleron process must be decoupled right after the electroweak symmetry breaking. 
This condition (the so-called sphaleron decoupling condition) is given by
\begin{align}
\Gamma^{(b)}_B(T) < H(T) \label{sphaleron},
\end{align}
where $\Gamma^{(b)}_B(T)$ is the baryon number changing rate in the broken phase, 
and $H(T)$ is the Hubble constant. Eq. (\ref{sphaleron}) can be translated into
\begin{align}
\frac{v(T)}{T}&>\frac{g_2}{4\pi{\cal{E}}(T)}\left[42.97+\log{\cal N}-2\log\left(\frac{T}{100~{\rm GeV}}\right)+\cdots\right] \nonumber \\
&\equiv \zeta_{\rm sph}(T) \label{v/T},
\end{align}
where the sphaleron energy is denoted as $E_{\rm sph}=4\pi v(T){\cal E}(T)/g_2$, 
with $g_2$ being the SU(2) gauge coupling. 
${\cal N}$ represents the translational and rotational zero-mode factors of the fluctuations 
about the sphaleron. 

In our numerical analysis, we first evaluate both $T_C$ and $v_C$,
where $T_C$ stands for a critical temperature at which the two degenerate minima appear 
in the effective potential, and $v_C$ is the VEV of the Higgs fields at $T_C$. 
The EWPT is studied in the direction of $\varphi$, and $\tan\beta$ is fixed by that at $T=0$.
We use the resummed finite-temperature one-loop effective potential
\begin{align}
V_{\rm eff}(\varphi,T)=&\sum_in_i
\Bigg[ \frac{\bar{M}^4_i(\varphi,T)}{64\pi^2}\left(\log\frac{\bar{M}^2_i(\varphi,T)}{\bar{\mu}^2}-c_i\right) \nonumber \\
&\hspace{1.5cm}+\frac{T^4}{2\pi^2}I_{B,F}\left(\frac{\bar{M}^2_i(\varphi,T)}{T^2}\right)\Bigg],
\end{align}
where
\begin{align}
I_{B,F}(a^2)=\int^{\infty}_0dx~x^2\log\left(1\mp e^{-\sqrt{x^2+a^2}}\right),
\end{align}
with the upper (lower) sign for bosons (fermions). $\bar{M}^2_i(\varphi,T)$ are the thermally corrected boson masses defined as $\bar{M}^2_i(\varphi,T)=\bar{m}^2_i(\varphi)+\Pi_i(T)$ where $\Pi_i(T)$ are the finite-temperature self-energy. 
Here, we consider the leading $\mathcal{O}(T^2)$ terms \cite{Carrington:1991hz}
\begin{align}
\Pi_\Phi(T) 
=&\frac{T^2}{12v^2}\Big[6m_W^2+3m_Z^2+m_H^2+m_A^2+2m_{H^\pm}^2 \nonumber\\
&\hspace{1cm}+6\big(m_t^2+m_b^2\big)\Big],
\\
\Pi_W(T) =& 2g_2^2T^2,\quad \Pi_B(T) = 2g_1^2T^2,
\end{align}
where $\Pi_\Phi$ for the Higgs bosons, $\Pi_W$ and $\Pi_B$ for the SU(2) and U(1) gauge bosons, respectively.
Note that the only longitudinal part of the gauge boson self-energy is thermally corrected. 

After finding $T_C$, we evaluate the sphaleron energy at that temperature 
(for a detailed calculation, see, e.g., Refs. \cite{Manton:1983nd,Klinkhamer:1984di,Funakubo:2009eg}).
Since the dominant contribution in the right-handed side of Eq. (\ref{v/T}) comes from ${\cal E}(T)$, we neglect  the logarithmic terms in our numerical analysis.
%
%
%
%
%
%
%
%
%
%
%
%
%
%
%
%
%
%
%
%
%
%
%
%
%
%
%
%

\indent
It should be noted that the EWPT does not start at $T=T_C$.
In order for the EWPT to occur, the bubbles must be nucleated at somewhat below $T_C$. 
Only bubble that has some critical size, which is called the critical bubble, can grow. 
The EWPT proceeds to develop if the bubble nucleation rate is larger than a certain value, 
and then the Universe is finally filled with the broken phase. 
We define the nucleation temperature ($T_N$) by the condition
\begin{align}
\Gamma_N(T_N)/H^3(T_N)=H(T_N),\label{defTn}
\end{align}
where $\Gamma_N(T_N)$ denotes the bubble nucleation rate per unit time per unit volume at $T_N$ \cite{Linde:1981zj}. 
From Eq.~(\ref{defTn}), it follows that
\begin{align}
\lefteqn{\frac{E_{\rm cb}(T_N)}{T_N}-\frac{3}{2}\log\frac{E_{\rm cb}(T_N)}{T_N}} \nonumber\\
&=152.59-2\log g_*(T_N)-4\log\left(\frac{T_N}{100~{\rm GeV}}\right), \label{cri_tem}
\end{align}
where $E_{\rm cb}(T_N)$ is the energy of the critical bubble 
and $g_*(T_N)$ represents the degrees of freedom of the relativistic particles at $T_N$.
As seen from Eq.~(\ref{cri_tem}), $E_{\rm cb}/T\lesssim150$ is needed for development of the EWPT.

We closely follow a method in~\cite{Funakubo:2009eg} to evaluate $E_{\rm cb}(T)$.
The critical bubbles are estimated from the following energy functional
\begin{align}
E_{\rm cb}(T)=&\int d^3\boldsymbol{x} \Big[ (\partial_i\Phi_1)^{\dagger}\partial_i\Phi_1 
+ (\partial_i\Phi_2)^{\dagger}\partial_i\Phi_2 \nonumber \\
&\hspace{3cm}+V_{\rm eff}(\Phi_1,\Phi_2,T) \Big],
\end{align}
where the gauge fields are assumed to take the pure-gauge configuration 
so that they do not contribute to the energy of the critical bubbles.
The classical Higgs fields are parametrized as
\begin{align}
\Phi_1(r)=\frac{1}{\sqrt{2}}
\begin{pmatrix}
0\\
\rho_1(r)
\end{pmatrix},\quad
\Phi_2(r)=\frac{1}{\sqrt{2}}
\begin{pmatrix}
0\\
\rho_2(r)
\end{pmatrix},
\end{align}
where $\rho_1(r)=\rho(r)\cos\beta$, $\rho_2(r)=\rho(r)\sin\beta$, $r=|\boldsymbol{x}|$, and here
$\tan\beta$ is fixed by that at $T=0$ as mentioned above.
In the numerical analysis, it is convenient to change $r$ and $\rho_i$
into the following dimensionless quantities:
\begin{align}
\xi=v(T)r,\quad h_1(\xi) = \frac{\rho_1(r)}{v(T)\cos\beta},\quad
	h_2(\xi) = \frac{\rho_2(r)}{v(T)\sin\beta}.
\end{align}
The profiles of $h_i(\xi)$ are obtained by solving the equations of motion
\begin{align}
-\frac{1}{\xi^2}\frac{d}{d\xi}\left(\xi^2\frac{dh_{1(2)}}{d\xi} \right)+\frac{1}{v^4(T)\cos^2\beta(\sin^2\beta)}\frac{dV_{\rm eff}}{dh_{1(2)}}=0,
\end{align}
with the boundary conditions: $\left. dh_{1,2}(\xi)/d\xi\right|_{\xi=0}=0$ and  $h_{1,2}(\xi=\infty)=0$. 
With those solutions, $E_{\rm cb}(T)$ is evaluated.

It is known that the bubble solutions are approximately given by a kink-configuration
\begin{align}
\rho_i(r)\sim v_i(T)\left[1-\tanh\left(\frac{r-R}{L_w}\right)\right], 
\end{align}
where $R$ and $L_w$ are the radius and wall width of the bubbles, respectively.
We use this as the initial configuration to derive the bubble solutions 
by using the relaxation method.
For more details about the numerical method, see, e.g., Ref.~\cite{Funakubo:2009eg}. 

%
%
%
%
\section{Numerical results}\label{sec:Results}
%
%
%
%
\begin{figure}[t]
\center
\includegraphics[width=8cm]{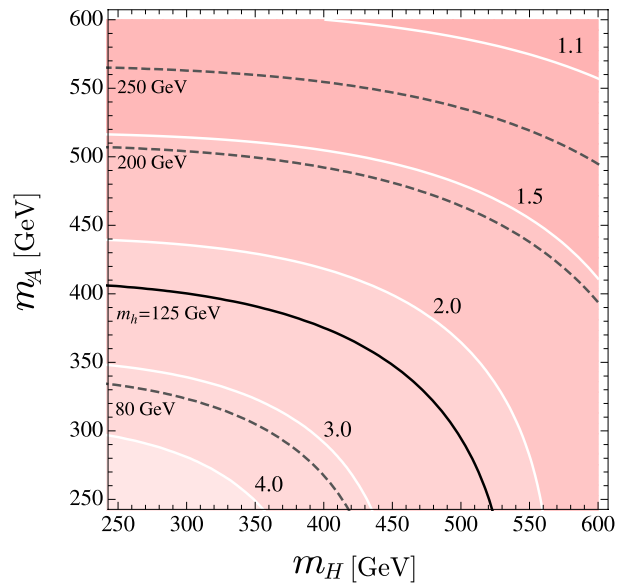} 
\caption{Contours of the Higgs boson mass and $v_C/T_C$ in the $(m_H,~m_A)$ plane. 
The solid line in black corresponds to $m_h=125$ GeV, 
and the dashed lines in gray indicate $m_h=80,~200$ and 250 GeV from bottom to top.
The each contour in white represents $v_C/T_C=1.1$, 1.5, 2.0, 3.0 and 4.0 from top to bottom.}
\label{fig:mH_vs_mA}
\end{figure}
%
%
%
%
In the SI-2HDM, there are five parameters in the tree-level potential:
\begin{align}
\lambda_1,\hspace{0.5cm}\lambda_2,\hspace{0.5cm}\lambda_3,\hspace{0.5cm}\lambda_4,\hspace{0.5cm}\lambda_5.
\end{align}
In our analysis, we replace them with the following physical parameters:
\begin{align}
m_H,\hspace{0.5cm}m_A,\hspace{0.5cm}m_{H^{\pm}},\hspace{0.5cm}\beta,\hspace{0.5cm}v.
\end{align}
We take $m_{H^\pm}=m_A$ as mentioned in section~\ref{sec:Model}.
Since $V_{\rm eff}$ does not depend on $\tan\beta$ explicitly, the results obtaining from it 
do not either, except for the cutoff of the model, as will be discussed in the following.

In Fig.~\ref{fig:mH_vs_mA}, we show the contours of the Higgs boson mass 
and $v_C/T_C$ in the $(m_H,~m_A)$ plane. 
The black solid (gray dashed) line indicates the parameter region 
where $m_h=$ 125 (80,~200,~250) GeV. 
As can be seen from Eq.~(\ref{Higgs_mass}), 
the Higgs boson mass gets larger as $m_H$ and $m_A$ increase. 
The white contours represent the magnitude of $v_C/T_C$. 
These contours indicate that the size of $v_C/T_C$ becomes smaller 
as $m_H$ and $m_A$ get heavier. 
Since the thermal effects from the heavy Higgs bosons 
cause the first-order EWPT in this model, 
$v_C/T_C$ would be proportional to $v\sum_im_i^3/\sum_im_i^4$, $i=H, A, H^{\pm}$, 
from the high-temperature approximation argument. 
This may explain the behavior of $v_C/T_C$ qualitatively. 
As a benchmark point, we take $m_h=125$ GeV and $m_H=m_A=382$ GeV. 
In this case, we find that $v_C/T_C=211~{\rm GeV}/91.5~{\rm GeV}=2.31$ 
and $\zeta_{\rm sph}(T_C)=1.23$.
Therefore, even though $\zeta_{\rm sph}$ is greater than the conventional criterion by 23\%,
$v_C/T_C$ is large enough to satisfy the sphaleron decoupling condition.
It is also found that $\zeta_{\rm sph}(T_C)$ is almost constant on the black line
while $v_C/T_C$ gets slightly weaken in the region where $m_H\gtrsim500$ GeV
since the thermal effect of $H$ loop is suppressed. 

\begin{figure}[t]
\center
\includegraphics[width=8cm]{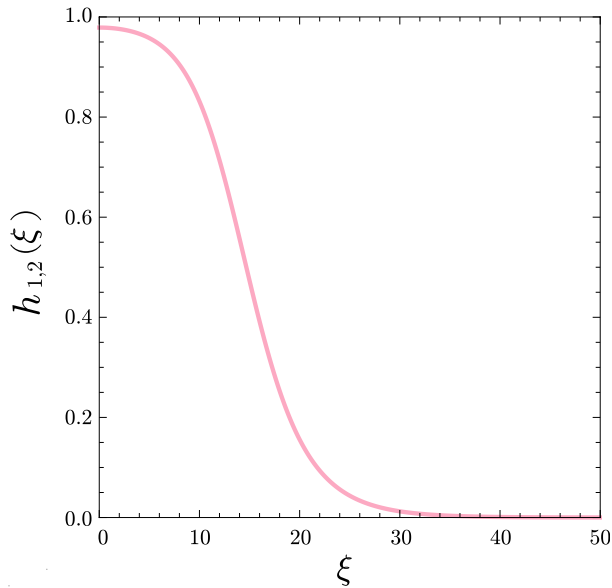} 
\caption{Bubble profiles of $h_{1,2}(\xi)$ at $T=T_N$. 
In this plot, we set $m_h=125$ GeV and $m_H=m_A=382$ GeV.}
\label{fig:bub_pro}
\end{figure}

In Fig.~\ref{fig:bub_pro}, the profiles of $h_{1,2}(\xi)$ is shown, here $h_1(\xi)=h_2(\xi)$ by construction.
Our numerical calculation shows that $v_N/T_N=229~{\rm GeV}/77.8~{\rm GeV}=2.94$,
$\zeta_{\rm sph}(T_N)=1.20$ and $E_{\rm cb}(T_N)/T_N=151.7$.
The degrees of the supercooling is about 15\%, i.e., $(T_C-T_N)/T_C=0.15$, 
which is more or less the same as the previous estimate~\cite{Funakubo:1993jg}.

Let us briefly make a comparison between the SI-2HDM and the minimal supersymmetric SM (MSSM).
Unlike the SI-2HDM, the supercooling in the MSSM case is rather small, e.g.,
 $\mathcal{O}(10^{-3})$ \cite{Funakubo:2009eg},
and the bubble wall width in the SI-2HDM is thinner than that in the MSSM, which is due to 
the stronger first-order EWPT compared to the MSSM case.
Since a CP violating source term may be proportional to the gradient of the bubble wall~\cite{ewbg}, 
the baryon number generation may be more efficient than the MSSM case. 
To this end, of course, the current model has to be extended to have an extra source of CP violation.

In Table~\ref{benchmark}, our numerical results in a benchmark point are summarized. 
In the SI-2HDM, the strong first-order EWPT 
is the inevitable consequence from the requirement of the 125 GeV Higgs boson.
In this case, the significant deviations may appear in $\mu_{\gamma\gamma}$ and $\lambda_{hhh}$.
We leave the study on the detectability of the heavy Higgs bosons to future work.

Finally, we comment on the cutoff scale ($\Lambda$) of this model.
Here, $\Lambda$ is determined by a scale at which $|\lambda_i| > 4\pi$ is obtained. 
In doing so, we use the one-loop renormalization group equations~\cite{Branco:2011iw}.
As an example, $\tan\beta=1$ is taken. 
It is found that $\Lambda = 6.3$ TeV, which is extremely low compared 
to a typical grand unification scale, $\sim 10^{16}$ GeV.
Since $\lambda_1\propto \tan^2\beta$ and $\lambda_2\propto 1/\tan^2\beta$,
the cases for $\tan\beta>1$ and $\tan\beta<1$ yield the lower cutoff scales than 6.3 TeV. Our analysis has reconfirmed the previous results \cite{Takenaga:1995ms, Hill:2014mqa}.
\begin{table}[t]
\center
\begin{tabular}{|c|c|}
\hline
$m_H$ & 382~GeV  \\
\hline
$v_C/T_C$ &  211~GeV/91.5~GeV~=~2.31 \\
$\zeta_{\rm sph}(T_C)$ & 1.23  \\
\hline
$v_N/T_N$ &  229~GeV/77.8~GeV~=~2.94 \\
$\zeta_{\rm sph}(T_N)$ & 1.20  \\
$E_{\rm cb}(T_N)/T_N$ & 151.7 \\
\hline
$\kappa_V$ & 1.0\\
$\kappa_f$ & 1.0\\
$\mu_{\gamma\gamma}$ & 0.90 \\
$\Delta\lambda_{hhh}$ & 82.1$\%$ \\
\hline
$\Lambda$ & 6.3~TeV \\
\hline
\end{tabular}
\caption{The benchmark point for the strong first-order EWPT and the nucleation of the bubbles, 
where we take $m_h=125$~GeV and $m_H=m_A=m_{H^{\pm}}$. 
For the evaluation of $\Lambda$, $\tan\beta=1$ is used.}
\label{benchmark}
\end{table}
%
%
%
%
\section{Conclusions and discussions}\label{sec:conclusion}
We have revisited the EWPT and the profiles of the critical bubbles in the SI-2HDM
in the light of the 125 GeV Higgs boson.
We improved these analyses by using the finite temperature one-loop effective potential 
with thermal resummation.
In this model, the heavy Higgs mass scales are fixed to be consistent with the $m_h=125$ GeV.
In our benchmark point, $m_H=m_A=m_{H^\pm}=382$ GeV, we found that 
$v_C/T_C=211~{\rm GeV}/91.5~{\rm GeV}=2.31$ and $\zeta_{\rm sph}(T_C)=1.23$. 
At the nucleation temperature, they are changed into $v_N/T_N = 229~{\rm GeV}/77.8~{\rm GeV}=2.94$ 
and $\zeta_{\rm sph}(T_N)=1.20$. 
Even though $\zeta_{\rm sph}$ in the SI-2HDM 
is greater than the conventional criterion by about 20\%, 
the first-order EWPT is strong enough to satisfy the sphaleron decoupling condition.

We also studied the deviations of the Higgs boson couplings from the SM predictions.  
It was found that even though the Higgs boson couplings to the gauge bosons and fermions 
are SM like, the significant deviations may appear 
in the $h\to \gamma\gamma$ mode and the triple Higgs boson coupling 
due to the nondecoupling effects of the heavy Higgs boson loops. 
In our benchmark point, the Higgs signal strength of $h\to\gamma\gamma$ is reduced by 10\%
and the triple Higgs boson coupling is enhanced by 82.1\%.
Such deviations may be detectable in the future experiments 
such as the High-Luminosity LHC~\cite{HL-LHC}
and the ILC~\cite{Baer:2013cma}.

There are some issues to be solved. 
In order to obtain the baryon asymmetry, an extra source of CP violation is needed as mentioned 
in the previous section.
To this end, we may augment this model by adding the extra fermions in a scale-invariant way.
The current analysis would not be much modified as long as the strength of the interactions between 
new particles and the Higgs boson are moderate.
Furthermore, since the cutoff of the model is rather small, the UV completion is needed.
However, constructing a complete model is beyond the scope of this Letter.

%
%
%
%
%
%
%
%
%
%
%
%
%
%
%
%
%
%
%
%
%
%
%
%
%
%
%
%
%
%
%
%
%
%
%
%
%
%
%
%
%
%
%
%
%
%
%
%
%
%
%
%
\begin{acknowledgments}
The work of K.F. is supported by Research Fellowships of the Japan Society for the Promotion of Science for Young Scientists. E.S. is supported in part by the Ministry of Science and Technology of R. O. C. under Grant No. MOST 104-2811-M-008-011.
\end{acknowledgments}


\end{document}